\documentclass[extendedabs]{recpad2k}

\usepackage{booktabs}

\title{User Specific Adaptation in Automatic Transcription of Vocalised Percussion} %MUDAR O NOME User Specific training in automatic transcription of vocalised percussion

\addauthor{Ant\'{o}nio Ramires}{antonio.ramires@inesctec.pt}{12}
\addauthor{Rui Penha}{rui.penha@inesctec.pt}{12}
\addauthor{Matthew E. P. Davies}{mdavies@inesctec.pt}{1}

\addinstitution{
 INESC TEC \\
 Sound and Music Computing Group\\
 Rua Dr. Roberto Frias, s/n, 4200 - 465 Porto, Portugal 
}\addinstitution{
 Faculty of Engineering\\
 University of Porto\\
 Rua Dr. Roberto Frias, s/n, 4200 - 465 Porto, Portugal 
}

\runninghead{Student, Prof, Collaborator}{RECPAD Author Guidelines}

\begin{document}

\maketitle

\sloppy

\begin{abstract}
The goal of this work is to develop an application that enables music producers to use their voice to create drum patterns when composing in Digital Audio Workstations (DAWs). An easy-to-use and user-oriented system capable of automatically transcribing vocalisations of percussion sounds, called LVT - Live Vocalised Transcription, is presented.\footnote{This work is derived from the MSc dissertation of Ant\'{o}nio Ramires, conducted in the Department of Electrical and Computer Engineering in the Faculty of Engineering, University of Porto.} LVT is developed as a Max for Live device which follows the ``segment-and-classify'' methodology for drum transcription, and includes three modules: i) an onset detector to segment events in time; ii) a module that extracts relevant features from the audio content; and iii) a machine-learning component that implements the k-Nearest Neighbours (kNN) algorithm for the classification of vocalised drum timbres.

Due to the wide differences in vocalisations from distinct users for the same drum sound, a user-specific approach to vocalised transcription is proposed. In this perspective, a given end-user trains the algorithm with their own vocalisations for each drum sound before inputting their desired pattern into the DAW. The user adaption is achieved via a new Max external which implements Sequential Forward Selection (SFS)  for choosing the most relevant features for a given set of input drum sounds. %and a new annotated dataset of vocalised drum sounds.

The evaluation of LVT addresses two objectives. First, to investigate the improvement in performance with user-specific training, and second, to assess if LVT can provide an optimised workflow for music production in Ableton Live when compared to existing drum transcription algorithms. Obtained results demonstrate that both objectives are met. 
\end{abstract}

%------------------------------------------------------------------------- 
\section{Introduction}
\label{sec:intro}
The development of computers' performance capacity, and the consequent possibility for real-time Digital Signal Processing (DSP) for audio, led to the appearance of Digital Audio Workstations (DAWs), making the creation of computer music available to the general public. Following these advances, many new instruments and interfaces for creating electronic music have surfaced. With changes in music culture, music production and how musicians work with their instruments has also changed. In other words, the ability to invent and reinvent the way to produce music is key to progress. Consequently, new proposals are necessary, such as designing new techniques for the composition of music. 
%While many such devices and DAWs exist, they typically require high technical expertise on the part of the user, and thus can create a barrier for novice users. 

Within the genre of Electronic Music, sequencing drum patterns plays a critical role. However, inputting drum patterns into DAWs often requires high technical skill on the part of the user, either by physically performing the patterns by tapping them on MIDI drum pads, or manually entering events via music editing software. For non-expert users both options can be very challenging, and can thus provide a barrier to entry. However, the voice is an important and powerful instrument of rhythm production, so it can be used to express or ``perform'' drum patterns in a very intuitive way - so called ``beatboxing.'' In order to leverage this concept within a computational system, our goal is towards a system to help users (both expert musicians and amateur enthusiasts) input rhythm patterns they have in mind into a sequencer via the automatic transcription of vocalised percussion. Our proposed tool is beneficial both from the perspective of workflow optimisation (by providing accurate real-time transcriptions), but also as means to encourage users to engage with technology in the pursuit of creative activities. From a technical standpoint, we seek to build on the state of the techniques from the domain of music information retrieval (MIR) for drum transcription \cite{matthew, gilletrichard} but actively targeted towards end-users and real-world music content production scenarios.   

%Therefore there is a compelling need to create systems which can leverage music understanding technology, in order to allow non-expert users to participate in music creative tasks. 

%In parallel to music creation software and hardware, the music information retrieval (MIR) community has been developing a wide range of techniques for the analysis and transcription of musical content. These MIR techniques hold the 

%While these MIR techniques have matured over the last 15 years, they are still subject to error. Thus    

%In particular, these seek to address the technical difficult and challenges for non-expert users in creating However there is still a high demand for new controllers as means to input or compose musical content. 

%The developments in music information retrieval (MIR) and in machine-learning paved the way for systems capable of transcribing drum loops and beatboxing. However, these systems are focused on evaluating the performance of transcription algorithms in offline testing scenarios and are either not easy to operate for end-users or not sufficiently reliable for use in a real music production workflow. 

\section{Methodology}
A vocalised drum transcription software, LVT, able to be trained with the user vocalisations is proposed. LVT is developed as a Max for Live project -- a visual programming environment, based on Max 7\footnote{www.cycling74.com}, which allows users to build instruments and effects for use within the Ableton Live\footnote{http://www.ableton.com/en/} DAW. 
To develop LVT, a dataset of vocalised percussion was compiled. A group of 20 participants (11 male, 9 female) were asked to record two short vocalised percussion tracks, one identical for all participants, and the other, an improvised pattern. These input percussion tracks were recorded three times: on a low quality laptop microphone, on an iPad microphone, and using a studio quality microphone (AKG c4000b). All recorded audio tracks were manually annotated using Sonic Visualiser\footnote{http://www.sonicvisualiser.org/}, a free application for viewing and analysing the contents of music audio files. The participants spanned a wide range of experience in beatboxing (from beatboxing experts, to those who had never vocalised drum patterns before), and covered a wide age range. Thus, we consider the annotated dataset to be representative of a wide range of potential users of the system, and highly heterogeneous in terms of the types of drum sounds.     

\begin{figure}[ht]
  \centering
  \includegraphics[width=.6\linewidth]{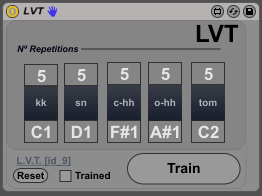}
  \caption{User interface of the LVT device.}
  \label{fig:1}
\end{figure}

Our proposed vocalised percussion transcription system was developed following a user-specific approach. LVT follows the ``segment and classify'' method for drum transcription \cite{gilletrichard} and integrates three main elements: i) an onset detector -- to identify when each drum sound occurs, ii) a component that extracts features for each event, and iii) a machine learning component to classify the drum sounds. In the Max for Live environment, the onset detection was performed with {\tt AubioOnset}$\sim$ \footnote{https://aubio.org/manpages/latest/aubioonset.1.html}. %using the high-frequency content detection function \cite{hfconset} - shown to be well suited for percussion sounds.
Feature extraction was performed in real-time using existing Max objects: {\tt Zsa.mfcc}$\sim$ -- to characterise the timbre, {\tt Zsa.descriptors} \cite{zsa} -- to provide spectral centroid, spread, slope, decrease and rolloff features \cite{zsa}, and finally the zero crossing rate and number of zero crossings were calculated with the {\tt zerox}$\sim$ object.
The machine learning component is trained with the user's preferred vocalisation and the features are selected which give the best results for the provided input. This is achieved using the Sequential Forward Selection method \cite{Whitney} along with a k-Nearest Neighbours classification algorithm, with the most significant features selected by the accuracy obtained from testing the training data (in our case, the annotated improvised patterns from each participant). SFS works by selecting the most significant feature, according to a specific parameter (in this case the classification accuracy), and adding it to an initially empty set until there are no improvements or no features remain. The k-NN algorithm was implemented using timbreID\cite{timbre}, and a new external for Max was developed to implement the SFS.
A user interface was created in Max for Live to facilitate the utilisation of the application by end-users. A screenshot of the interface of LVT is shown in Fig. \ref{fig:1}. It demonstrates the user-specific training stage -- where a user inputs a set number of the drum timbres they intend to use, after which their vocalised percussion is transcribed and rendered as a MIDI file for subsequent synthesis. 

%The system is composed by two parts, LVT and LVT receiver. In Max for Live, to convert audio to Midi, two devices are necessary: an audio effect, LVT, that processes the audio and, then, broadcast a message to LVT receiver, a midi effect that, when a message is received, creates MIDI notes in Ableton Live.
%The first step when using this system is to set the number of  training repetitions for each vocalisation. If one of the percussive instruments is not required, setting this value to 0 will disable it. Next, the desired MIDI notes for each vocalisation can be set. These should be the notes that correspond to the desired sounds in the instrument that follows the LVT receiver in the MIDI chain. To train the system, the ``Train'' button should be pressed. The user should then start vocalising the desired drum sounds by repeating each sound the number of times defined earlier. Once the system has been trained, it will then automatically recognise the different drum sounds produced by the user and transcribe them to a MIDI clip.

To operate LVT, a user loads the device in Ableton Live and then vocalises the set of desired drum sounds they intend to use, e.g. five kick sounds followed by five snare sounds, followed by five hi-hat sounds. Once the expected number of drum sounds have been detected, the SFS algorithm then identifies the subset of features which best separate the drum sounds for the user. After training, the user can then vocalise rhythmic patterns which are automatically converted from audio to a MIDI representation in the DAW for later synthesis and editing.     

\section{Results}
The evaluation of LVT was designed to serve two purposes. First, to understand how a user-specific trained system performs against state of the art drum transcription system (which have been optimised over large datasets without any user-specific training), and second, to explore how LVT could improve a producer's workflow. We compared LVT against two existing drum transcription algorithms: LDT \cite{matthew}, and Ableton Live's built-in ``Convert Drums to MIDI'' function. For validation data we used the non-improvised vocalised patterns from our annotated dataset.   

To compare the accuracy of the systems we use the F-measure of the transcriptions. Then, to investigate how our system could improve a producers workflow, the ``effort'' to get an accurate transcription was calculated by counting the number of editing operations required to obtain the desired patterns. These operations are as follows: to \emph{modify}, to \emph{add}, or to \emph{remove} a MIDI note. 

\begin{table}[t]
\centering 
\caption{Number of operations and F-measure for the AKG microphone.}
\label{tab:1}
\begin{tabular}{l c c c c c c}
\toprule 
     & \multicolumn{3}{c}{\it Edit Operations} & \multicolumn{3}{c}{\it F-measure}  \\

    & Modify & Add & Remove  & Kick & Snare & Hi-hat \\

    \midrule
    Ableton & 33 & 12 & 296 &  0.518 & 0.470 & 0.297\\
    LDT & 52 & 24 & 206 & 0.538 & 0.204 & 0.419\\ 
    LVT & 39 & 7 & 15  & 0.914 & 0.691 & 0.802  \\
    \bottomrule
\end{tabular}
\end{table}

Table \ref{tab:1} summarises the results obtained from counting the total number of operations needed to obtain the desired pattern for the testing data recorded on the studio quality AKG c4000b microphone and the corresponding F-measure per vocalised drum sound, on the three drum transcription systems. The results demonstrate that, for the studio quality microphone, vocalised drum transcription accuracy for LVT is substantially higher than the other systems, and far fewer modifications were required to obtain the desired patterns when editing the automatic transcriptions.

% \begin{table}[ht]
% \label{tab:1}
% \caption{Number of operations and F-measure for the AKG microphone.}
% \begin{tabular}{|l|ccc|ccc|}
% System & Modify & Add & Remove & Kick & Snare  & Hi-hat \\
% \hline
% LVT & 39 & 7 & 15  & 0.914 & 0.691 & 0.802  \\
% Ableton & 33 & 12 & 296 &  0.518 & 0.470 & 0.297\\
% LDT & 52 & 24 & 206 & 0.538 & 0.204 & 0.419\\ \hline

% %if space left split the table

% \end{tabular}
% \end{table}

\begin{figure}[t!]
  \begin{center}
    \leavevmode
    \includegraphics[width=.9\linewidth]{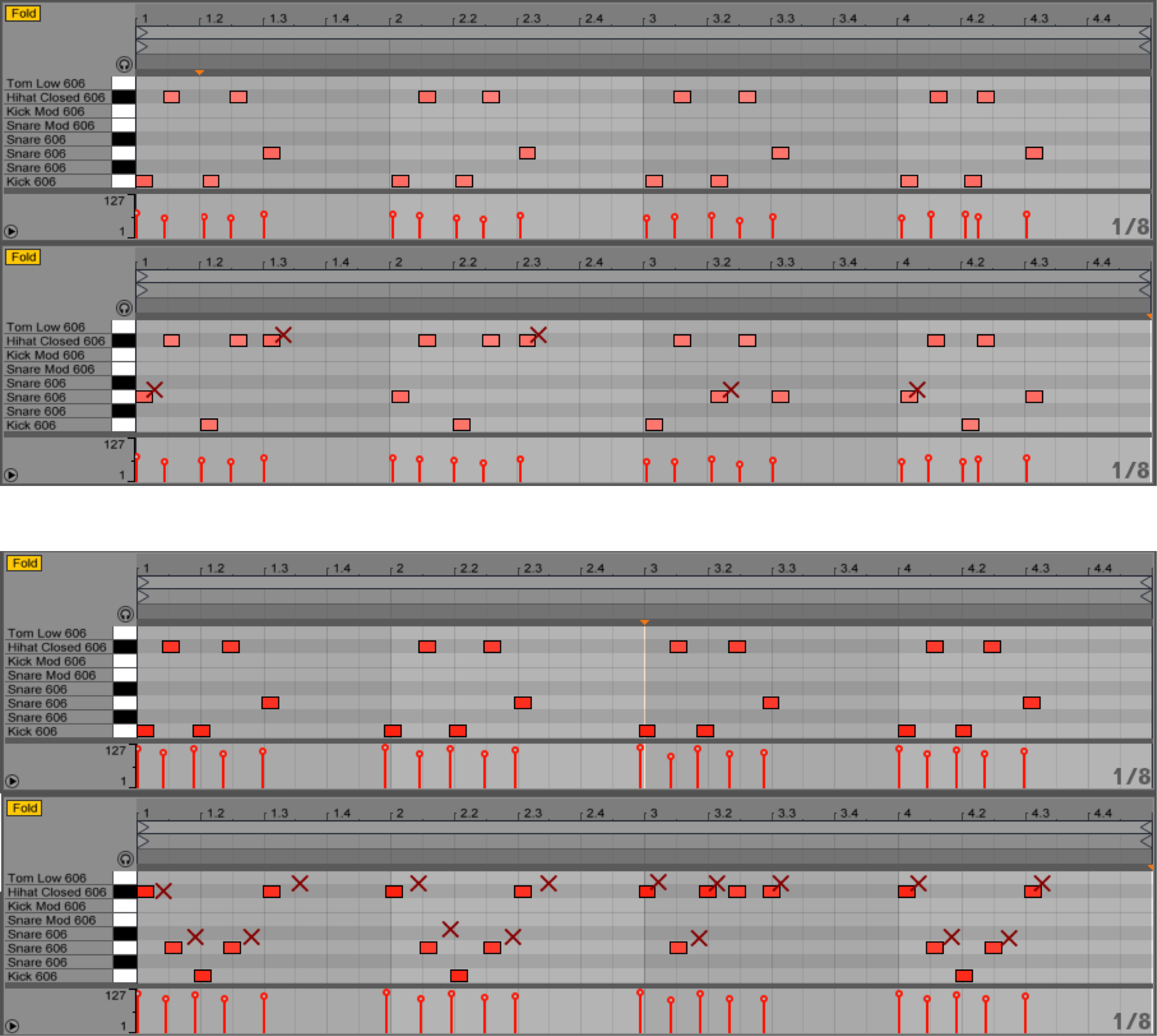}
    \caption{(top) First user vocalisations trained with the second user. (bottom) Second user vocalisations trained with the first user.}
    \label{fig:combined}
  \end{center}
\end{figure}

To see the effect of user-specific training on the performance of LVT, an example is provided where LVT is trained on one user and tested on another -- and vice-versa. When training the LVT with a different person with different vocalisations, the accuracy of the transcription is decreased as shown in Fig. \ref{fig:combined}. In the upper part of each screenshot is the transcription of the user when trained with its own vocalisations, while the bottom part corresponds to the transcription when trained with the other user. As can be seen, without the user-specific training, many misclassifications occur.

%The timing measurements for each system to present a transcription of one element of the dataset to the user is approximately 12.9s for the Ableton Live system, 13.1s for LVT and 6.9s for LDT.

By examining the previously obtained results, we infer that LVT can provide a transcription closer to the ground truth than the existing state of the art systems, as shown by the higher F-measure. In addition to LVT being trained per individual user, these results may also derive from the fact that LVT does not try to detect polyphonic events (more than one drum vocalisation at the same time) as the other systems do. Furthermore, LVT does not detect as many events as the other systems, and this has a strong influence on the number of false positives, and hence the F-measure. The number of events to achieve the desired transcription, presented in Table \ref{tab:1}, shows that the end-user of the system does not have to perform as many actions when producing music, which has a positive impact on the workflow, leaving more time for creative experimentation.

%exapmple of user training
%\begin{figure}[t!]
%  \begin{center}
%    \leavevmode
%    \includegraphics[width=.85\linewidth]{treinoISApatJSIL2}
%    \caption{First user vocalisations trained with the second user.}
%    \label{fig:isajsil}
%  \end{center}
%\end{figure}

%\begin{figure}[t!]
%  \begin{center}
%    \leavevmode
%    \includegraphics[width=.85\linewidth]{TreinoJSILpatISA2}
%    \caption{Second user vocalisations trained with the first user.}
%    \label{fig:jsilisa}
%  \end{center}
%\end{figure}

\section{Conclusions}

In this paper, we have presented LVT -- a new interface for assistive music content creation. LVT allows Ableton Live users to sequence MIDI patterns that can be used for designing and performing rhythms with their voice. Existing state of the art systems, including one already in Ableton Live, are not able to transcribe vocalised percussion as effectively because these tools are trained for general recorded drum sounds which are typically not vocalised. Indeed, because different people vocalise drum sounds in different ways, LVT explicitly seeks to model and capture this behaviour via user-specific training. Our evaluation shows LVT to be very effective for wide range of users and vocalisations, outperforming existing systems. Furthermore, we believe LVT can be applied to any kinds of arbitrary non-pitched percussive sounds -- provided that the training sound types are sufficiently different from one another, and can thus be well separated in the audio feature space using SFS.
%As each user can choose the desired vocalisations for each drum sound, the system is versatile enough to also transcribe real drums or any kind of non-pitched percussive sounds. As long as the training sounds are different enough from each other, LVT is able to automatically choose the features that provide a good separation and therefore good classification accuracy for any input via the use of the SFS feature selection method. 

LVT is implemented as a Max for Live device, and thus fully integrates into Ableton Live, allowing users of all ability ranges to experiment with music sequencing driven by their own personal percussion vocalisations within an easy-to-use graphical user interface.

%------------------------------------------------------------------------- 
\section{Acknowledgements}
This work is financed by the ERDF - European Regional Development Fund through the Operational Programme for Competitiveness and Internationalisation - COMPETE 2020 Programme within project \guillemotleft POCI-01-0145-FEDER-006961\guillemotright, and by National Funds through the FCT - Funda\c{c}\~{a}o para a Ci\^{e}ncia e a Tecnologia (Portuguese Foundation for Science and Technology) as part of project  UID/EEA/50014/2013.

Project TEC4Growth-Pervasive Intelligence, Enhancers and Proofs of Concept with Industrial Impact/NORTE-01-0145-FEDER-000020 is financed by the North Portugal Regional Operational Programme (NORTE 2020), under the PORTUGAL 2020 Partnership Agreement, and through the European Regional Development Fund (ERDF). 
%------------------------------------------------------------------------- 

\bibliography{egbib}

\end{document}